\documentclass[fleqn,twoside]{article}
\usepackage{espcrc2}
\usepackage{amsmath,amstext,amsfonts,amsbsy,amssymb,amscd,bbm,epsfig,graphicx,subfigure}
\usepackage{graphicx}
\usepackage[figuresright]{rotating}

\renewcommand{\AmS}{{\protect\the\textfont2
  A\kern-.1667em\lower.5ex\hbox{M}\kern-.125emS}}
\newcommand{\ba}{\begin{array}}
\newcommand{\ea}{\end{array}}

\newcommand{\rep}[1]{\cite{#1}}
\newcommand{\refig}[1]{Fig.~\ref{#1}}
\newcommand{\ret}[1]{Table~\ref{#1}}

\newcommand{\Dslash}{\relax{\kern+.25em / \kern-.70em D}}

\newcommand{\fm}{\rm fm}

\newcommand{\GeV}{\rm GeV}

\newcommand{\Real}{\relax{\mathsf{\Gamma\kern-.35em R}}}
\newcommand{\Int}{\relax{\mathsf{Z\kern-.40em Z}}}

\newcommand{\SUt}{\mbox{SU}(3)}

\newcommand{\MSbar}{{\overline{\rm MS}}}

\newcommand{\gbar}{\kern1pt\overline{\kern-1pt g\kern-0pt}\kern1pt}
\newcommand{\mbar}{\kern2pt\overline{\kern-1pt m\kern-1pt}\kern1pt}
\newcommand{\obar}[1]{\kern3pt\overline{\kern-2pt #1\kern-0pt}\kern1pt}

\newcommand{\Oa}{\mbox{O}(a)}
\newcommand{\Oasq}{\mbox{O}(a^2)}

\newcommand{\abar}{\kern1pt\overline{\kern-1pt a\kern-0pt}\kern1pt}

\newdimen\arrayruleHwidth
\makeatletter
\def\Hline{\noalign{\ifnum0=`}\fi\hrule \@height \arrayruleHwidth
  \futurelet \@tempa\@xhline}
\makeatother
\title{\vspace{-5.0cm}
       \rightline{\normalsize ROM2F/2004/25}
       \vspace{-0.1cm}
       \rightline{\normalsize MS-TP-04-23}
       \vspace{-0.1cm}
       \rightline{\normalsize DESY 04-145}
       \vspace{-0.1cm}
       \rightline{\normalsize FTUAM-04-20}
       \vspace{-0.1cm}
       \rightline{\normalsize IFT-UAM/CSIC-04-47}
       \vspace{-0.1cm}
       \rightline{\normalsize September 2004}
       \vspace{2.0cm}
       Precision computation of $B_K$ in quenched
       lattice QCD\thanks{Talk presented by C.~Pena at Lattice 2004, Fermilab.}
        \setcounter{footnote}{0}
      }

\author{P.~Dimopoulos\address[ToV]{INFN Sezione di Roma II,
        c/o Dipartimento di Fisica, Universit\`a di Roma ``Tor Vergata'',\\
        \ Via della Ricerca Scientifica 1, I-00133 Rome, Italy},
        J.~Heitger\address[WWUM]{Westf\"alische Wilhelms-Universit\"at
        M\"unster, Institut f\"ur Theoretische Physik,\\
        ~\,Wilhelm-Klemm-Strasse 9, D-48149 M\"unster, Germany},
        C.~Pena\address[DESY-HH]{DESY, Theory Group,
        Notkestrasse 85, D-22603 Hamburg, Germany},
        S.~Sint\address[UAM]{Departamento de F\'{\i}sica Te\'orica
        C-XI and Instituto de F\'{\i}sica Te\'orica C-XVI,\\
        ~\,Universidad Aut\'onoma de Madrid,
        Cantoblanco, E-28049 Madrid, Spain}
        and
        A.~Vladikas\addressmark[ToV]
        (ALPHA Collaboration)
       }

\begin{document}

\begin{abstract}
%\vspace{1pc}
We present the results of a precision computation of $B_K$ with Wilson fermions.  
Simulations are performed at different lattice spacings, enabling continuum limit
extrapolations. Two different twisted mass QCD (tmQCD) regularisations are considered for the
computation of bare matrix elements. In both cases the relevant four-fermion operator
renormalises multiplicatively. In one regularisation it is possible to perform the
computation directly at the physical kaon mass value, thus avoiding extrapolations in the
mass. Nonperturbative renormalisation is carried out using available Schr\"odinger Functional
results.
\end{abstract}

\maketitle

\section{tmQCD for $B_K$}

The kaon mixing parameter $B_K$ is currently the dominant source of
uncertainty in unitarity triangle analyses of CP violation. It is
therefore of utmost importance to have as precise a lattice QCD
determination of this quantity as possible. Our project aims at a
computation of $B_K$ with Wilson fermions that brings all the
systematics (apart from quenching) under control, in order to obtain a final
result with an uncertainty of at most a few percent.

The main sources of uncertainty in existing quenched computations with
Wilson fermions are the limits to simulations imposed by the presence
of exceptional configurations and the mixing under renormalisation of
the parity-even part of the relevant four-fermion operator
\begin{gather}
O^{\Delta S=2} = (\bar s_{\rm L}\gamma_\mu d_{\rm L})
                 (\bar s_{\rm L}\gamma_\mu d_{\rm L}) \ .
\end{gather}
Both
problems can be eliminated with the use of tmQCD Wilson
regularisations.\footnote{For a review of tmQCD and recent
  developments related to it, see~\rep{Frezzotti_plenary}.}
A framework to compute $B_K$ without the need of
determining mixing coefficients, via a tmQCD regularisation with a fully
twisted $(u,d)$ doublet and an untwisted $s$ quark, was introduced
in~\rep{Frezzotti:2000nk}. The first results obtained within this approach have been
presented in previous Lattice conferences~\rep{BK_procs}. Here we present our
first analysis of the full set of results, and supply a preliminary
value for $B_K$ in the continuum limit.

We always work within a Schr\"odinger Functional (SF) framework. Our action is
nonperturbatively $\Oa$ improved in the bulk of the SF cylinder, and
one-loop $\Oa$ improved at the time boundaries. For the
nonperturbative renormalisation of the four-fermion operator we use
the results in~\rep{Guagnelli:2002rw}.

\section{Results}

Simulations have been performed at three values of the lattice
coupling $\beta$ and a number of different quark masses in the
limit of unbroken $\SUt_V$ symmetry $M_s=M_d$, where $M_i$ are the
physical renormalised quark masses. The parameters
employed in the runs from which our final value for $B_K$ is derived
are summarised in~\ret{tab:run_parameters}, together with the values
of the pseudoscalar meson mass and the bare value of $B_K$ obtained
in each case. The latter is extracted from a suitable ratio of correlation
functions, as described in~\rep{BK_procs}.

To obtain the value of $B_K$ at the physical point we
have extrapolated the results in~\ret{tab:run_parameters} to $(r_0
M_{\rm PS})^2=1.591$, assuming a linear dependence of $B_K$ on $M_{\rm PS}^2$
(which actually leads to our best fits). We stress that an independent
simulation has been performed at each value of the mass, and
therefore the results are fully uncorrelated.

\begin{table*}[!ht]
 \centering
 \begin{tabular}{c@{\hspace{5mm}}c@{\hspace{3mm}}c@{\hspace{5mm}}lll@{\hspace{7mm}}ll}
 \Hline\\[-1.5ex]
 $\beta$ & lattice & \# config. & ~~~$\kappa_s$ & ~~~~$\kappa_d$ & ~~~$a\mu_d$ & ~~$aM_{\rm PS}$ & ~~~$B_K$ \\[1.0ex]
 \hline\\[-2.0ex]
 $6.0$ & $16^3 \times 48$ & 402 & $0.1335$ & $0.135169$  & $0.03186$   & $0.3895(11)$ & $1.025(8)$ \\
       &                  & 398 & $0.1338$ & $0.135178$  & $0.03152$   & $0.3551(12)$ & $0.993(11)$ \\
       &                  & 402 & $0.1340$ & $0.135183$  & $0.02708$   & $0.3320(11)$ & $0.977(11)$ \\
       &                  & 400 & $0.1342$ & $0.135187$  & $0.02261$   & $0.3054(11)$ & $0.943(10)$ \\[0.5ex]
\hline\\[-2.0ex]
 $6.2$ & $24^3 \times 64$ & 200 & $0.1346$ & $0.135780$  & $0.028324$  & $0.2815(8)$  & $0.980(7)$ \\
       &                  & 201 & $0.1347$ & $0.135783$  & $0.025985$  & $0.2685(9)$  & $0.952(8)$ \\
       &                  & 214 & $0.1349$ & $0.135787$  & $0.021290$  & $0.2432(10)$ & $0.934(8)$ \\[0.5ex]
\hline\\[-2.0ex]
 $6.3$ & $24^3 \times 72$ & 204 & $0.1348$ & $0.135771$  & $0.023639$  & $0.2385(10)$ & $0.970(14)$ \\
       &                  & 212 & $0.1349$ & $0.135773$  & $0.021254$  & $0.2251(11)$ & $0.937(12)$ \\
       &                  & 205 & $0.1351$ & $0.135776$  & $0.016467$  & $0.1964(12)$ & $0.885(15)$ \\[1.0ex]
 \Hline\\[0.0ex]
 \end{tabular}
 \caption{Parameters for the runs from which our main data have been
 extracted. The bare twisted $d$ quark mass
 has been denoted by $\mu_d$. The mass parameters are tuned
 so as to have equal renormalised $s$ and $d$ quark masses up to
 $\Oasq$ cutoff effects.}
 \label{tab:run_parameters}
\end{table*}

\begin{figure}[t!]
\vspace{3.0cm}
\includegraphics{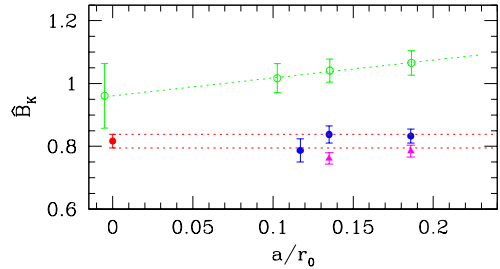}
\caption{
Extrapolation (solid points, band) to the continuum limit of the renormalisation group
invariant $B$-parameter $\hat B_K$. To allow for an easy comparison
with a reference computation employing Wilson fermions, we include the
results of the method without subtractions of~\rep{Becirevic:2004aj} (empty points),
which display a linear extrapolation in $a/r_0$. Also displayed
(triangles) are the results for $B_K$ obtained with $(\pi/4)$-twisted
quarks (see text).
}
\vspace{-0truemm}
\label{fig:CL_extrap}
\end{figure}

After nonperturbative renormalisation, we can extrapolate our results
to the continuum limit as depicted in~\refig{fig:CL_extrap}. Since the
matrix element is not fully $\Oa$ improved, the renormalisation group
invariant parameter $\hat B_K$ is expected
to approach the continuum limit linearly in $a/r_0$; however, we
observe no cutoff dependence within the statistical accuracy of the
results, and therefore take a fit to a constant as our best
result. As a preliminary value in the continuum limit we finally quote
\begin{gather}
  \begin{split}
    ~~&\hat B_K = 0.817(22)\ , \\
    ~~&\obar{B}_K^{\MSbar,\rm{NDR}}(2~\GeV) = 0.592(16) \ .
  \end{split}
\end{gather}

\begin{figure}[t!]
\vspace{3.0cm}
\includegraphics{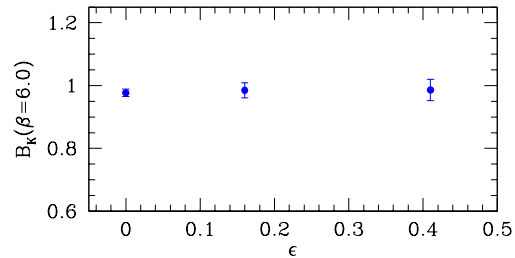}
\caption{
Dependence of the bare $B_K$ at $\beta=6.0$ on the $\SUt_V$ breaking parameter $\epsilon$.
}
\vspace{-0truemm}
\label{fig:eps_dependence}
\end{figure}

\begin{figure}[t!]
\vspace{3.0cm}
\includegraphics{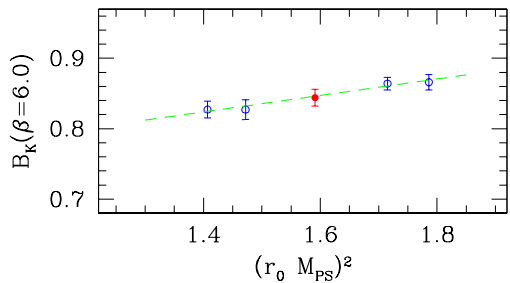}
\caption{
Interpolation to obtain $B_K(\beta=6.0)$ at the physical
kaon mass in the $(\pi/4)$-twisting approach (solid point) from the
results for four different pseudoscalar masses (open points).
}
\vspace{-0truemm}
\label{fig:mass_interp}
\end{figure}

We have performed a number of tests to ensure that the quoted error
adequately takes into account a number of systematic uncertainties.
First of all, simulations on larger lattices (corresponding to
$L\approx 2.2~\fm$) at $\beta=6.0$ have shown that the value of $B_K$
at the lightest pseudoscalar meson mass does not depend on the
physical volume within its statistical uncertainty. We also performed
a study, working at $\beta=6.0$ and $r_0 M_{\rm PS} \simeq
1.777$, of the dependence of $B_K$ on the breaking of the $\SUt_V$
symmetry, parameterised by
$\epsilon=(M_s-M_d)/(M_s+M_d)$. \refig{fig:eps_dependence} shows the
results: no significant dependence is observed up to
$\epsilon \sim 0.4$.

\section{Eliminating mass extrapolations}

The main remaining source of systematic uncertainty in the previous
method comes from the fact that using an untwisted $s$ quark prevents
us from computing at physical values of the kaon mass in the $M_s=M_d$
limit. There are various tmQCD regularisations that allow to avoid
this problem. One simple possibility consists in considering a
regularisation in which the doublet $(s,d)$ is twisted with angle
$\alpha=\pi/4$. It can be easily shown that the renormalisation
properties of the relevant four-fermion operator are exactly the same
as in the previous case. This approach
is designed for mass degenerate strange and down quark masses, and,
since both quarks are twisted, the physical kaon mass can now be reached
without any problem. An
alternative that uses only fully twisted ($\alpha=\pi/2$) quarks has
been proposed in~\rep{Frezzotti:2004wz}. An attractive feature of this
latter proposal is that it is expected to bring in automatic $\Oa$
improvement of the (bare) matrix element.

As a feasibility study of the $(\pi/4)$-twisting approach we have performed
simulations at $\beta=6.0,6.2$ and a number of different
pseudoscalar meson masses, on a box of physical length $L \approx
1.5~\fm$. The value of $B_K$ at the physical kaon point is then
obtained by linear interpolation in
$M_{\rm PS}^2$, as shown e.g. in~\refig{fig:mass_interp}.
The results are also shown in~\refig{fig:CL_extrap}.
They exhibit remarkably small uncertainties,
and suggest a good agreement with the outcome from the previous
approach. However, preliminary results with the same run parameters
on a larger physical volume hint at the presence of a small but,
given the small statistical uncertainties, still relevant finite
volume effect. This issue has therefore to be settled by more
extensive simulations.

\section*{Acknowledgements}
\noindent Work supported in part by the European Union's Human
Potential Programme under contract HPRN-CT-2000-00145, Hadrons/Lattice
QCD.


\begin{thebibliography}{9}

\bibitem{Frezzotti_plenary}
R.~Frezzotti,
%``Wilson fermions with chirally twisted mass,''
Nucl.\ Phys.\ Proc.\ Suppl.\  {\bf 119} (2003) 140;
%[arXiv:hep-lat/0210007].
%%CITATION = HEP-LAT 0210007;%%
these proceedings.

\bibitem{Frezzotti:2000nk}
%R.~Frezzotti, P.~A.~Grassi, S.~Sint and P.~Weisz  [ALPHA collaboration],
R.~Frezzotti et al. [ALPHA Collab.],
%``Lattice QCD with a chirally twisted mass term,''
JHEP {\bf 0108} (2001) 058.
%[arXiv:hep-lat/0101001].
%%CITATION = HEP-LAT 0101001;%%

\bibitem{BK_procs}
%M.~Guagnelli, J.~Heitger, C.~Pena, S.~Sint and A.~Vladikas,
M.~Guagnelli et al. [ALPHA Collab.],
%``K0 anti-K0 mixing from the Schroedinger functional and twisted mass  QCD,''
Nucl.\ Phys.\ Proc.\ Suppl.\  {\bf 106} (2002) 320;
%[arXiv:hep-lat/0110097].
%%CITATION = HEP-LAT 0110097;%%
%P.~Dimopoulos, J.~Heitger, C.~Pena, S.~Sint and A.~Vladikas [ALPHA Collaboration],
P.~Dimopoulos et al. [ALPHA Collab.],
%``B(K) from twisted mass QCD,''
Nucl.\ Phys.\ Proc.\ Suppl.\  {\bf 129} (2004) 308.
%[arXiv:hep-lat/0309134].
%%CITATION = HEP-LAT 0309134;%%

\bibitem{Guagnelli:2002rw}
%M.~Guagnelli, J.~Heitger, C.~Pena, S.~Sint and A.~Vladikas [ALPHA Collaboration],
M.~Guagnelli et al. [ALPHA Collab.],
%``Non-perturbative scale evolution of four-fermion operators,''
Nucl.\ Phys.\ Proc.\ Suppl.\  {\bf 119} (2003) 436; to appear.
%[arXiv:hep-lat/0209046].
%%CITATION = HEP-LAT 0209046;%%

\bibitem{Becirevic:2004aj}
%D.~Be\'cirevi\'c, P.~Boucaud, V.~Gim\'enez, V.~Lubicz and M.~Papinutto,
D.~Be\'cirevi\'c et al.,
%``B(K) from the lattice with Wilson quarks,''
arXiv:hep-lat/0407004.
%%CITATION = HEP-LAT 0407004;%%

\bibitem{Frezzotti:2004wz}
R.~Frezzotti and G.C.~Rossi,
%``Chirally improving Wilson fermions. II: Four-quark operators,''
arXiv:hep-lat/0407002.
%%CITATION = HEP-LAT 0407002;%%

\end{thebibliography}
\end{document}